\pdfoutput=1
\documentclass[a4paper,11pt]{article}
\usepackage[utf8]{inputenc}

\usepackage[table]{xcolor}
\usepackage{jcappub}
\usepackage{graphicx}
\usepackage{psfrag,fancyhdr,epsfig}
\usepackage{hyperref}
\hypersetup{colorlinks,bookmarksopen,bookmarksnumbered,citecolor=magenta,linkcolor=blue,pdfstartview=FitH,urlcolor=blue}

\usepackage{amsmath}
\usepackage{array}
\usepackage{graphicx}
\usepackage{color}
\usepackage{tensor}
\usepackage{xcolor}
\usepackage{orcidlink}

\newcommand{\be}{\begin{equation}}
\newcommand{\ee}{\end{equation}}
\newcommand{\bear}{\begin{array}}
\newcommand{\eear}{\end{array}}
\newcommand{\ba}{\begin{eqnarray}}
\newcommand{\ea}{\end{eqnarray}}
\newcommand{\phimine}{\phi_\text{min, $\epsilon$}}
\newcommand{\phimaxe}{\phi_\text{max, $\epsilon$}}
\newcommand{\phimaxk}{\phi_\text{max, $k$}}

\def\a{\alpha}
\def\b{\beta}

\def\d{\delta}
\def\e{\epsilon}
\def\eps{\epsilon}
\def\g{\gamma}

\def\l{\lambda}
\def\m{\mu}
\def\n{\nu}

\def\tns{\tensor}
\def\cR{\mathcal{R}}

\usepackage[most]{tcolorbox}

\tcbset{colback=yellow!10!white, colframe=red!50!black, 
        highlight math style= {enhanced, %<-- needed for the ’remember’ options
            colframe=red,colback=red!10!white,boxsep=0pt}
        }

        \usepackage{empheq}
\newtcbox{\mymath}[1][]{%
    nobeforeafter, math upper, tcbox raise base,
    enhanced, colframe=blue!30!black,
    colback=blue!30, boxrule=1pt,
    #1}

\title{Symmetry-breaking inflation in non-minimal metric-affine gravity}

\author[]{Ioannis D.~Gialamas\orcidlink{0000-0002-2957-5276}}
\author[]{and Antonio Racioppi\orcidlink{0000-0003-4825-0941}}

\emailAdd{ioannis.gialamas@kbfi.ee}
\emailAdd{antonio.racioppi@kbfi.ee}

\affiliation[]{Laboratory of High Energy and Computational Physics, 
National Institute of Chemical Physics and Biophysics, R{\"a}vala pst.~10, Tallinn, 10143, Estonia}

\abstract{We study symmetry-breaking inflation within the framework of metric-affine gravity. By introducing a non-minimal coupling, $\beta(\phi)\tilde{\cal R}$, between the Holst invariant and the inflaton, both small-field and large-field inflationary predictions can be brought into agreement with the latest observational constraints. Remarkably, even for sub-Planckian vacuum expectation values, appropriately chosen values of $\beta(\phi)$ enable viable inflation—a scenario previously considered unattainable.}

\begin{document}

\maketitle

\section{Introduction}
\label{introduction}

Cosmological inflation provides a compelling framework to explain the observed structure and features of the Universe~\cite{Kazanas:1980tx,Sato:1980yn,Guth:1980zm,Linde:1981mu}. During this phase of rapid, quasi-de Sitter expansion, quantum fluctuations in gravitational and matter fields were amplified into the cosmological perturbations~\cite{Starobinsky:1979ty,Mukhanov:1981xt,Hawking:1982cz,Starobinsky:1982ee,Guth:1982ec,Bardeen:1983qw} responsible for the large-scale structures we observe today. The process is commonly modeled using a scalar field, known as the inflaton, which generates the required vacuum energy and seeds spatial inhomogeneities. Observational evidence, including data from the cosmic microwave background (CMB), large-scale structures, and supernovae, supports a flat, homogeneous, and isotropic Universe. Furthermore, quantum mechanical treatment of inflation offers a robust mechanism for producing primordial anisotropies. However, recent analyses, such as those from the Planck mission~\cite{Planck:2018jri}, have imposed stringent constraints on inflationary models, ruling out many simplistic scenarios, such as the symmetry-breaking inflation (SBI) model examined in this study.

Initially SBI has been introduced as an explicit realization of the \emph{new inflation} scenario \cite{Albrecht:1984qt,Moss:1985wn}. Later it has also been studied as a completion of the hilltop inflationary model (e.g. \cite{Boubekeur:2005zm,Lillepalu:2022knx} and refs. there in). Additional SBI raises in the context of gauge mediated supersymmetry breaking scenarios \cite{Dine:1997kf,Riotto:1997iv}, where the underlying supersymmetry is protecting the inflaton potential from radiative corrections that might spoil the flatness of the potential itself. Later on also the possibility of non-minimal couplings to gravity have been investigated (e.g. \cite{Accetta:1985du,Belfiglio:2023rxb}) but only in the case of metric gravity. The case of the metric-affine realizations lied unstudied, hence the purpose of our work.

In particular, we move beyond the standard metric formulation of General Relativity, where the connection is the Levi-Civita connection, to explore how inflationary observables in the SBI model are modified within the metric-affine formulation, where the connection is treated as an independent variable. While the two formulations are equivalent under the standard Einstein-Hilbert action, they diverge when modifications such as non-minimal couplings of scalar fields to curvature or quadratic curvature terms are introduced.
In the context of metric-affine gravity (MAG), an additional scalar quantity, that is linear in the Riemann tensor, can be formed, alongside the standard Ricci scalar. This is the so-called Host invariant~\cite{Hojman:1980kv,Nelson:1980ph,Holst:1995pc}, which is the contraction of the Riemann tensor with the antisymmetric Levi-Civita tensor, and it can lead to interesting phenomenological implications~\cite{BeltranJimenez:2019hrm,Langvik:2020nrs,Shaposhnikov:2020aen,Shaposhnikov:2020gts,Iosifidis:2020dck,Piani:2022gon,Pradisi:2022nmh,Salvio:2022suk,Gialamas:2022xtt,DiMarco:2023ncs,He:2024wqv,Barker:2024dhb,Gialamas:2024jeb,Racioppi:2024zva,Karananas:2024xja,Inagaki:2024ltt,Capozziello:2024lsz,Gialamas:2024iyu,Racioppi:2024pno}.  As we will see later, general MAG theories involve non-zero torsion, which is crucial for the presence of the Holst invariant in the action. Additionally, while non-metricity is also present, it does not play a significant role in our case.

The paper is structured as follows: In Section~\ref{the_model}, we introduce the SBI model with a non-minimal coupling to gravity within the framework of MAG. Section~\ref{sec:pot_kin} discusses the behavior of the kinetic function and the scalar field potential in detail. In Section~\ref{sec:inf}, we thoroughly analyze the inflationary predictions of the SBI model. Finally, our conclusions are presented in Section~\ref{sec:concl}.

\section{The model}
\label{the_model}

We start with the Jordan frame action:
\be 
S_{\rm JF}= \int {\rm d}^4x\sqrt{-g}\left[\alpha(\phi){\cal R}+\beta(\phi)\tilde{\cal R}   - \frac12 (\partial_\mu \phi)^2 - V(\phi) \right], 
\label{eq:Sstart} 
\ee
where  $V(\phi)$ is the inflaton potential, and $\alpha(\phi)$ and $\beta(\phi)$ are non-minimal coupling functions. ${\cal R}$ and $\tilde{\cal R}$  denote, respectively, the scalar and pseudoscalar contractions of the curvature tensor (the latter also known as the Holst invariant~\cite{Hojman:1980kv,Nelson:1980ph,Holst:1995pc}) and are given by
\be
\cR = \tns{\cR}{^\a^\b_\a_\b}\,, \qquad \tilde{\cR} = g_{\a\m} \epsilon^{\m\b\g\d} \tensor{\cR}{^\a_\b_\g_\d}\,,
\ee
where $\epsilon^{\mu\nu\rho\sigma}$ is the Levi-Civita tensor\footnote{The Levi-Civita tensor is written as $\eps_{\a\b\g\d} =\sqrt{-g} \mathring{\varepsilon}_{\a\b\g\d}$, where $ \mathring{\varepsilon}_{\a\b\g\d}$ is the Levi-Civita symbol with $ \mathring{\varepsilon}_{0123}=1$. Note that the components of the Levi-Civita symbol with upper indices (i.e. $\mathring{\varepsilon}^{\a\b\g\d}$) are equal to the components of ${\rm sign}(g)\mathring{\varepsilon}_{\a\b\g\d} = -\mathring{\varepsilon}_{\a\b\g\d}$.} and  $\tensor{\cR}{^\a_\b_\g_\d}$ is the curvature associated with the metric-independent connection $\tns{\Gamma}{^\l_\m_\n}$, given by
\be
\label{eq:riem}
\tns{\cR}{^\a_\b_\g_\d} =\partial_\g \tns{\Gamma}{^\a_\d_\b} - \partial_\d \tns{\Gamma}{^\a_\g_\b} +\tns{\Gamma}{^\a_\g_\m}\tns{\Gamma}{^\m_\d_\b} -\tns{\Gamma}{^\a_\d_\m}\tns{\Gamma}{^\m_\g_\b}  \ .
\ee
To maintain minimality, we do not include in the action~\eqref{eq:Sstart}  terms that are nonlinear in the Riemann tensor~\eqref{eq:riem}. This ensures that the model describes only the massless graviton and the inflaton as physical degrees of freedom, avoiding terms that introduce higher-order derivatives.
To be more specific, adding quadratic terms in the Ricci scalar and/or the Holst invariant to the action~\eqref{eq:Sstart} would introduce an additional pseudoscalar degree of freedom. These models have been extensively studied in~\cite{Pradisi:2022nmh,Gialamas:2022xtt}. Additionally, terms constructed directly from the Ricci tensor would also result in new degrees of freedom.

We remind that in MAG, the connection  $\tns{\Gamma}{^\l_\m_\n}$ is not assumed to be the Levi-Civita one, $\{\tns{}{^\l_\m_\n}\}$. Instead, it is determined dynamically by its equation of motion. We also remind that, if $\tns{\Gamma}{^\l_\m_\n}$ is the Levi-Civita connection,  $\tilde{\cal R}$ vanishes\footnote{The careful reader might notice that $\tilde{\cal R}$ actually vanishes for any  $\tns{\Gamma}{^\l_\m_\n}$ that is symmetric in its last two indices.} and ${\cal R}$ equals to the standard Ricci scalar $R$ constructed from the Levi-Civita connection. MAG theories generally involve nonzero torsion, $T_{\l\m\n}=\mathcal{C}_{\l[\m\n]} \neq 0$, and are not metric-compatible, i.e., $Q_{\l\m\n} =\nabla_\l g_{\m\n} = -2\mathcal{C}_{(\n|\l|\m)}\neq 0$, where $\tns{\mathcal{C}}{^\l_\m_\n}$ is the distortion tensor, defined as the difference $\tns{\Gamma}{^\l_\m_\n}-\{\tns{}{^\l_\m_\n}\}$. Theories with zero torsion correspond to the Palatini case, while those with zero non-metricity belong to the Einstein-Cartan case. In the Palatini case, the Holst invariant is identically zero, meaning torsion vanishes. However, in our case, torsion is nonzero. On the other hand, non-metricity can be set to zero without loss of generality due to a projective symmetry of the action, $\tns{\mathcal{C}}{_\l_\m_\n} \rightarrow \tns{\mathcal{C}}{_\l_\m_\n} +g_{\l\n}A_\m $. Therefore, our action~\eqref{eq:Sstart} is dynamically equivalent to the Einstein-Cartan framework. This equivalence would not hold if other invariants directly constructed from the non-metricity, $Q_{\l\m\n}$, were included in the action.

In addition to the Ricci and Holst terms shown in action~\eqref{eq:Sstart}, MAG allows the construction of $20$ additional scalar quantities with mass dimension $2$, derived from torsion and non-metricity~\cite{Iosifidis:2021bad,Rigouzzo:2022yan}. Although these terms have the same dimension as the Ricci and Holst terms, we omit them, assuming that the Riemann curvature~\eqref{eq:riem} is the fundamental object.

Expressing the curvature scalars $\cR$ and $\tilde{\cR}$ in terms of the distortion tensor, we proceed to integrate out the non-dynamical distortion tensor. This process becomes straightforward if we vary the action with respect to it. The distortion obeys an algebraic, non-homogeneous linear equation of motion, and thus, to fully integrate it out, it is sufficient to find a particular solution~\cite{Pradisi:2022nmh,Gialamas:2022xtt,Gialamas:2023emn} (see also~\cite{Iosifidis:2021bad} for exact solutions for the affine-connection in MAG). After some algebra, we obtain
\be
\label{eq:EF_act}
\mathcal{S} = \int {\rm d}^4x \sqrt{-\bar{g}} \left[\frac{M_P^2}{2}R -\frac{k(\phi)}{2} (\partial_\mu \phi)^2 -U(\phi)  \right]\,,
\ee
where $M_P$ is the reduced Planck mass, with
\be
k(\phi) =  \frac{M_P^2}{2} \left[\frac{1}{\a} + \frac{12(\a'\b -\a \b')^2}{\a^2(\a^2+4\b^2)} \right]\,,  \qquad U(\phi)=\frac{V(\phi)}{F^2(\phi)} \, , \qquad
F(\phi) = \frac{2\a}{M_P^2}\,, \label{eq:k:U:F}
\ee
and the prime denotes derivative with respect the scalar field. 
Note that after substituting the distortion back into the action, the action is in the Jordan frame, and a Weyl rescaling of the form $g_{\m\n} = F^{-1}(\phi) \bar{g}_{\m\n}$ has been applied to obtain the Einstein frame action~\eqref{eq:EF_act}.

As mentioned before we are interested in non-minimally coupled SBI therefore
\be
 V(\phi) = \frac{\lambda}{4} \left(\phi^2 - v^2\right)^2 \, , \label{eq:SB:V}
\ee
where $\lambda$ is a real positive parameter and $v$ is the vacuum expectation value (vev) of the inflaton.
For what concerns the non-minimal coupling functions, we consider the most natural choice
\be 
\alpha(\phi) = \frac{M_P^2}{2} \left( \delta_\alpha^2 + \xi \frac{\phi^2}{M_P^2} \right)   , \qquad 
\beta(\phi) = \frac{M_P^2}{2} \left( \delta_{\beta}^2+ \tilde\xi \frac{\phi^2}{M_P^2} \right) \, .
\ee 
The form of the nonminimal coupling term $\a(\phi)$ is chosen primarily for renormalizability reasons. In quantum field theory, interactions should preserve renormalizability to ensure that divergences can be consistently controlled. The coupling constant 
$\xi$ is dimensionless, which ensures that the nonminimal coupling term does not introduce any new mass scales that could lead to nonrenormalizable divergences. This makes it the simplest and most natural way to couple a scalar field to gravity while maintaining the renormalizability of the theory. Additionally, such a coupling naturally arises as a counterterm in curved spacetime quantum field theory, reinforcing its necessity in a consistent effective field theory framework. Similar arguments apply to the choice of $\b(\phi)$. 
Note that all the functions involved in eq.~\eqref{eq:k:U:F} are symmetric under $\phi \to - \phi$, therefore, without loss of generality, form now on we work focus on $\phi>0$. The usual consistency constraint $\alpha(\phi)>0$ forces $\xi \geq 0$. On the other hand, $\beta(\phi)$ can be negative and therefore also $\tilde\xi$ can.

The toy models with $\xi >0$ and $\beta(\phi) = 0$ has been previously studied in e.g. \cite{Accetta:1985du,Belfiglio:2023rxb}. However, according to our knowledge, nobody has ever studied before the setup with $\xi=0$ and $\beta(\phi) \neq 0$. This is the purpose of our work. 
To conclude this Section, without loss of generality we also assume as customary $ \delta_\alpha =1$, which means that the theory is already set in the Einstein frame. Thus, no Weyl rescaling is needed, and only source of non-minimality is the coupling function $\beta$ to the Holst invariant (i.e. the kinetic function $k$).

\section{Study of the scalar potential and kinetic function}
\label{sec:pot_kin}

\begin{figure}[t]
\centering
\includegraphics[width=0.5\textwidth]{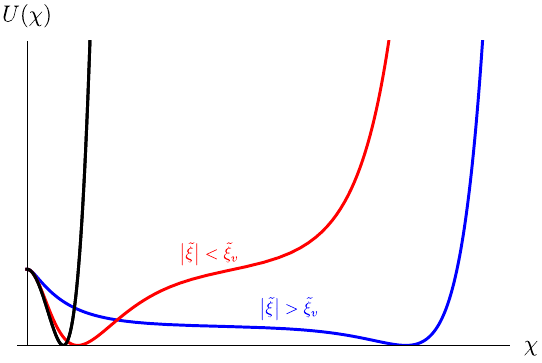}
\caption{$U(\chi)$ vs. $\chi$ for $\beta(\phi) = 0$ (black) and $\beta(\phi) \neq 0$ with $|\tilde\xi| > \tilde\xi_v $ (blue) and $|\tilde\xi| < \tilde\xi_v $ (red).}
\label{fig:Uplot}
\end{figure}

We begin by reminding some key concepts of SBI.  The inflationary potential is given in eq. \eqref{eq:SB:V} and a generic plot is presented in Fig. \ref{fig:Uplot} in black line. The potential exhibits a local maximum in $\phi=0$ and a global minimum in $\phi=v$. Two regions are available for inflation. First we have a small field region for $\phi < v$, where inflation is hilltop-like. Moreover, there is also a large field region for $\phi > v$. In this region the behavior gets closer to a quartic potential the smaller the vev compared to the Planck mass. In both cases, inflation ends when $\phi$ approaches $v$.

The presence of a kinetic function does not change this overall picture, but, as we will prove in the following, induces an additional flat region before or after $v$. After imposing $\alpha = \frac{M_P^2}{2}$, the kinetic function in eq. \eqref{eq:k:U:F} becomes
\be
\label{eq:kinfun}
k(\phi) = 1+ \frac{24\tilde{\xi}^2\phi^2 M_P^2}{M_P^4+4\left(\delta_{\beta}^2 M_P^2+\tilde{\xi}\phi^2\right)^2}\,.
\ee
Such a kinetic function has been already studied in \cite{Racioppi:2024pno} where it was proven that induces a new flat region of the potential via an inflection point. Such an inflection point can be tracked back to $k(\phi)$ exhibiting a local maximum in
\be
 \phimaxk^2 =  M_P^2 \frac{\sqrt{1+4 \delta _{\beta }^4}}{2 \left| \tilde{\xi }\right| } \, , \quad \text{with} \quad k(\phimaxk)=1+6 |\tilde\xi|  \left(\sqrt{4 \delta _{\beta }^4+1}-\text{sign}(\tilde\xi) 2 \delta _{\beta }^2\right) \label{eq:phi:max}
\ee
when $\tilde\xi$ is not null. Nearby such a maximum the flattening of the potential is enhanced. On the other hand, in the limits $\tilde{\xi}\phi^2 \gg M_P^2, \phi \gg M_P$, or where $\tilde{\xi}\phi^2 \ll M_P^2, \phi \ll M_P$, the standard case, i.e. $k(\phi) =1$, is recovered. We also stress that, at a given $\delta_\beta$, the value of $\tilde\xi$ controls whether the maximum of $k(\phi)$ is before or after $v$. Hence, it is convenient to compute the value of $\tilde\xi$ so that $\phi_\text{max, $k$}=v$
\be
 \tilde\xi_v =   \frac{M_P^2}{v^2} \frac{\sqrt{1+4 \delta _{\beta }^4}}{2 } \, .\label{eq:tilde:xi:v}
\ee
Therefore, if $|\tilde \xi|< \tilde\xi_v (> \tilde\xi_v)$ then $\phi_\text{max, $k$} > v (< v)$. In Figure~\ref{fig:Uplot} we show naively how the potential~\eqref{eq:SB:V}, for the canonical normalized scalar\footnote{The canonical normalized scalar $\chi$ is defined by solving the differential equation $ {\rm d}\chi = \sqrt{k(\phi)} {\rm d}\phi$. Due to the complexity of the kinetic function~\eqref{eq:kinfun}, this equation can only be solved numerically.} $\chi$, changes in both cases. A more detailed numerical analysis follows in the next Section.

\section{Inflationary results}
\label{sec:inf}
In this Section we discuss the inflationary predictions of our model. 
We analyze cosmological observables constrained by recent observations~\cite{Planck:2018jri,BICEP:2021xfz}, focusing initially on the amplitude of the primordial curvature perturbation generated on super-Hubble scales during slow-roll inflation. This amplitude is expressed as:
\be
A_s^\star = H_\star^2/(8\pi^2M_P^2\eps_1^\star)\,, \label{eq:As}
\ee
and it has been constrained to $A_s^\star \simeq 2.1\times 10^{-9}$ at the pivot scale $k_\star = 0.05 $ Mpc$^{-1}$~\cite{Planck:2018jri}.

Two key observables, the tensor-to-scalar ratio $r$ and the spectral index $n_s$ of the scalar power spectrum, are defined as:
\be
r=16\e_1\,\qquad \text{and} \qquad n_s = 1-2\eps_1-\eps_2\,,
\ee
while the running of the latter is given by
\be
\alpha_s= -2\e_1\e_2-\e_2\e_3\,.
\ee
Here, $\e_i$ are the Hubble flow functions, or slow-roll parameters, defined as:
\be
\e_1 = -{\rm d}\ln H/{\rm d} N\,, \quad \e_2 = {\rm d}\ln\eps_1/{\rm d} N\,, \quad  \e_3 = {\rm d}\ln\e_2/{\rm d} N\,, \quad {\rm d}N =H{\rm d}t\,.
\ee
Given the complexity of the kinetic function~\eqref{eq:kinfun}, analytic approximations of the observables are challenging to derive. Therefore, we rely on numerical solutions to the equations of motion under the slow-roll approximation, bypassing the need to solve the Mukhanov-Sasaki equation. In our numerical computations, we use the Hubble flow functions instead of the potential slow-roll parameters commonly employed in similar studies, {in order to get more precise results}. However, the potential slow-roll parameters
\be
 \epsilon_U = \frac{M_P^2}{2k(\phi)} \left(\frac{U'(\phi)}{U(\phi)}\right)^2 \qquad \text{and} \qquad
 \eta_U =M_P^2\frac{\left(k^{-1/2}(\phi)U'(\phi) \right)'}{k^{1/2}(\phi)U(\phi)}\,, \label{eq:eps:eta}
\ee
will still be useful as an immediate tool to interpret our results.

The evolution of the scalar field during inflation is closely tied to the number of $e$-folds, denoted by $N_\star$. Assuming instantaneous reheating after inflation, $N_\star$ is given by~\cite{Liddle:2003as}
\be
\label{eq:efolds}
N_\star = 66.5 - \ln \left[k_\star/(a_0 H_0)\right] +\frac14\ln\left[9H_\star^4/(\rho_{\rm end})\right]\,,
\ee
where the subscripts ``0'' and ``end'' refer to quantities evaluated at the present day and at the end of inflation, respectively. $\rho_{\rm end} = 3U(H_{\rm end})/2$ is the energy density of the inflaton at the end of inflation. The inflationary phase is followed by oscillations of the scalar field around the minimum of a quadratic-like potential, which provides a natural transition to the reheating stage. The inclusion of the matter sector is essential for a complete understanding of reheating dynamics. Moreover, the distinction between metric-affine gravity and other gravitational formulations becomes manifest in this context, potentially leading to distinct phenomenological consequences that warrant further investigation which fall beyond the purpose of the present work. 

As customary in SBI we divide our discussion in two subcases: small field and large field inflation. In the following, all the predictions regarding small field inflation will be shown with a dashed line, while the ones about large field inflation will be with a continuous one.

\subsection{Small field inflation}

%%%%%%%%%%%%%%%%%%%%%%%%%%FIGURE%%%%%%%%%%%%%%%%%%%%%%%%%%%%%%%%
\begin{figure}[t]
\centering
\includegraphics[width=0.49\textwidth]{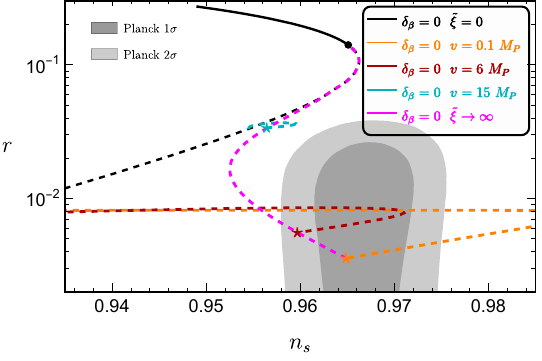}\,
\includegraphics[width=0.49\textwidth]{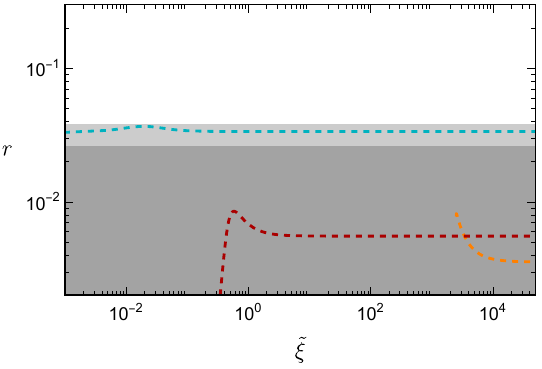}
\\
\includegraphics[width=0.49\textwidth]{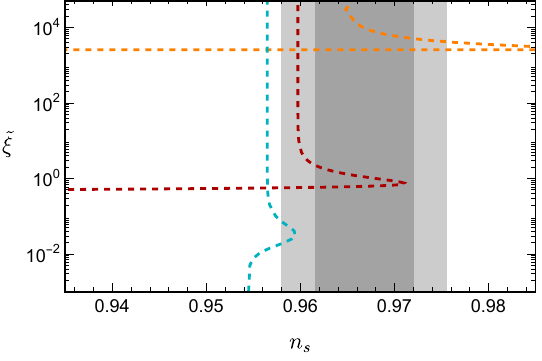}\,
\includegraphics[width=0.49\textwidth]{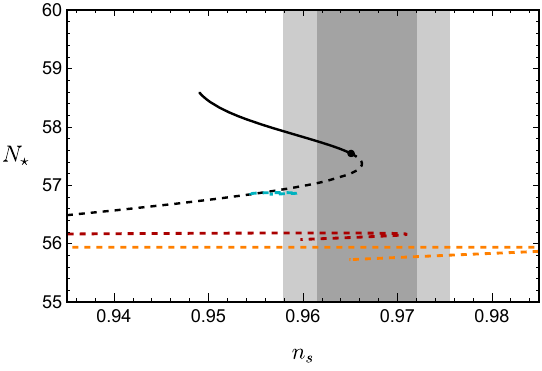}
\\
\includegraphics[width=0.49\textwidth]{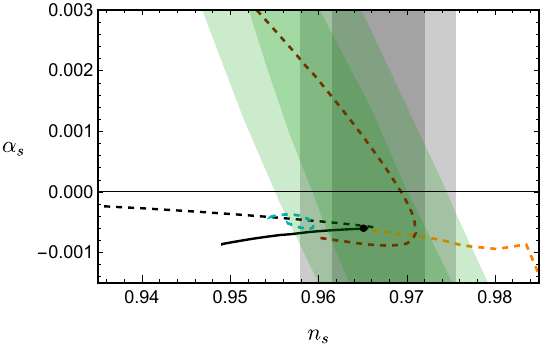}\,
\includegraphics[width=0.49\textwidth]{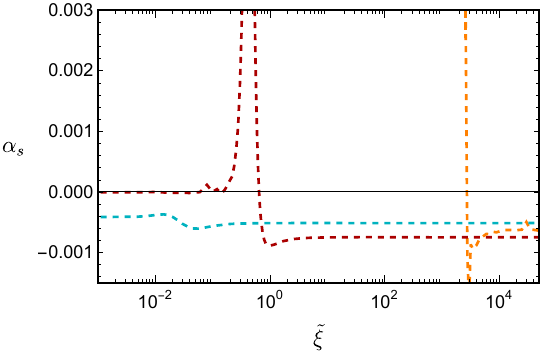}
\caption{Small field inflationary predictions for $\delta_\b=0$.
$r$ vs.~$n_\text{s}$ (up, left),  $r$ vs.~$|\tilde\xi|$ (up, right),  $|\tilde\xi|$ vs.~$n_\text{s}$ (center, left),  $N_*$ vs.~$n_\text{s}$ (center, right),  $\alpha_\text{s}$ vs.~$n_\text{s}$ (down, left) and $\alpha_\text{s}$ vs.~$|\tilde\xi|$ (down, right) for $N_\star$ given by~\eqref{eq:efolds} and $v=0.1 M_P$ (orange), $v=6 M_P$ (red) and $v=15 M_P$ (cyan). For reference, the predictions for $\tilde\xi=0$ (black) and $|\tilde\xi| \to \infty$ (pink). The gray (green) areas represent the 1,2$\sigma$ allowed regions coming  from  the latest combination of Planck, BICEP/Keck and BAO data~\cite{BICEP:2021xfz} (from Planck legacy data \cite{Planck:2018vyg}).}
\label{fig:inflation:small:beta0}
\end{figure}
%%%%%%%%%%%%%%%%%%%%%%%%%FIGURE_END%%%%%%%%%%%%%%%%%%%%%%%%%%%%%%%%%

%%%%%%%%%%%%%%%%%%%%%%%%%%FIGURE%%%%%%%%%%%%%%%%%%%%%%%%%%%%%%%%
\begin{figure}[t]
\centering
\includegraphics[width=0.49\textwidth]{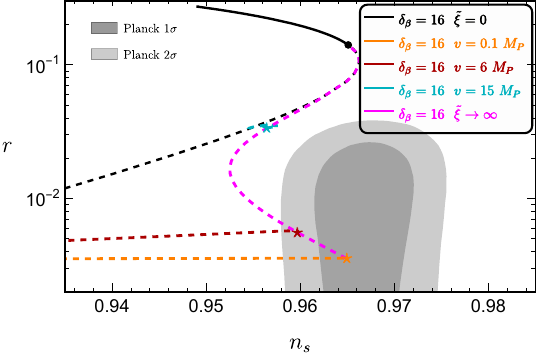}\,
\includegraphics[width=0.49\textwidth]{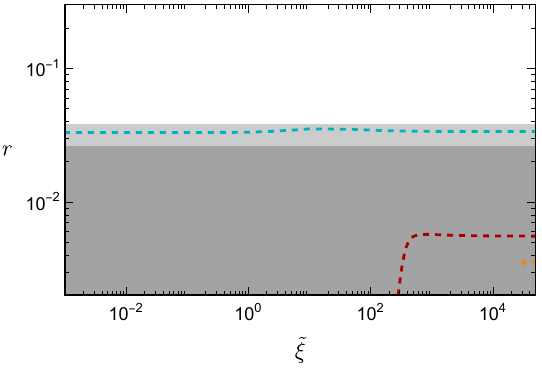}
\\
\includegraphics[width=0.49\textwidth]{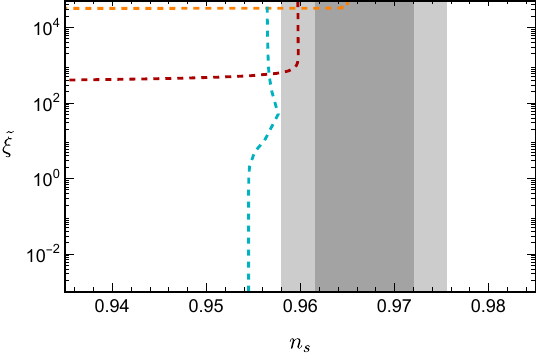}\,
\includegraphics[width=0.49\textwidth]{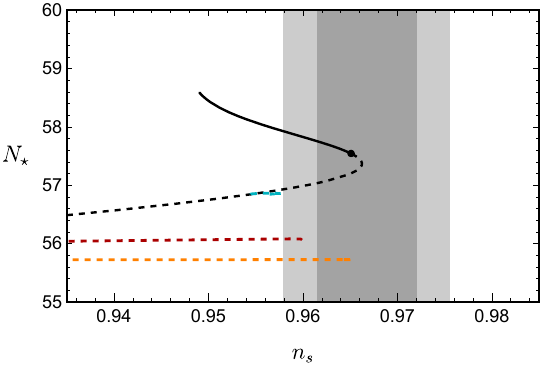}
\\
\includegraphics[width=0.49\textwidth]{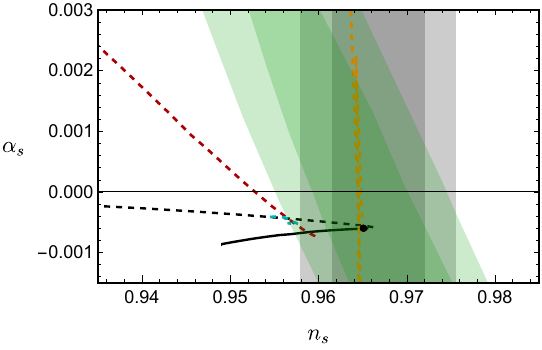}\,
\includegraphics[width=0.49\textwidth]{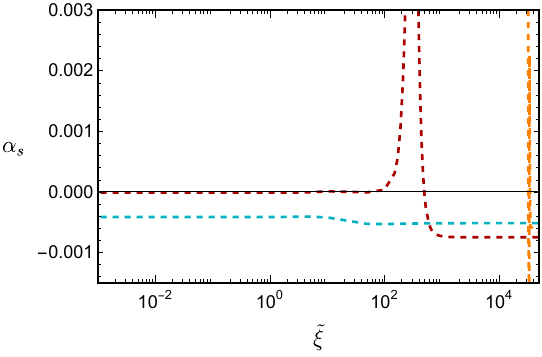}
\caption{Small field inflationary predictions for $\delta_\b=16$ and $\tilde{\xi}>0$. 
$r$ vs.~$n_\text{s}$ (up, left),  $r$ vs.~$|\tilde\xi|$ (up, right),  $|\tilde\xi|$ vs.~$n_\text{s}$ (center, left),  $N_*$ vs.~$n_\text{s}$ (center, right),  $\alpha_\text{s}$ vs.~$n_\text{s}$ (down, left) and $\alpha_\text{s}$ vs.~$|\tilde\xi|$ (down, right) for $N_\star$ given by and $v=0.1M_P$ (orange), $v=6M_P$ (red) and $v=15M_P$ (cyan). The black, pink, gray and green color code are the same as in Fig. \ref{fig:inflation:small:beta0}.}
\label{fig:inflation:small:beta16:pos}
\end{figure}
%%%%%%%%%%%%%%%%%%%%%%%%%FIGURE_END%%%%%%%%%%%%%%%%%%%%%%%%%%%%%%%%%

%%%%%%%%%%%%%%%%%%%%%%%%%%FIGURE%%%%%%%%%%%%%%%%%%%%%%%%%%%%%%%%
\begin{figure}[t]
\centering
\includegraphics[width=0.49\textwidth]{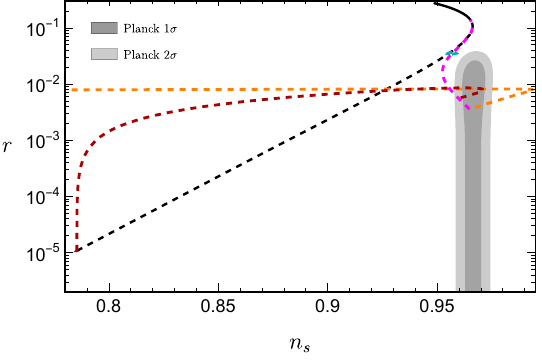}
\includegraphics[width=0.49\textwidth]{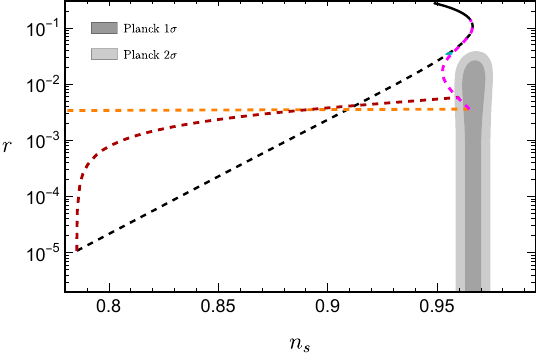}
\caption{$r$ vs. $n_s$ zoom out for small field inflation with $\tilde{\xi}>0$ with $\delta_\b=0$ (left) and $\delta_\b=16$ (right). The color code is the same as in Figs.~\ref{fig:inflation:small:beta0} and~\ref{fig:inflation:small:beta16:pos}.}
\label{fig:inflation:small:zoomout}
\end{figure}
%%%%%%%%%%%%%%%%%%%%%%%%%FIGURE_END%%%%%%%%%%%%%%%%%%%%%%%%%%%%%%%%%

%%%%%%%%%%%%%%%%%%%%%%%%%%FIGURE%%%%%%%%%%%%%%%%%%%%%%%%%%%%%%%%
\begin{figure}[t]
\centering
\includegraphics[width=0.49\textwidth]{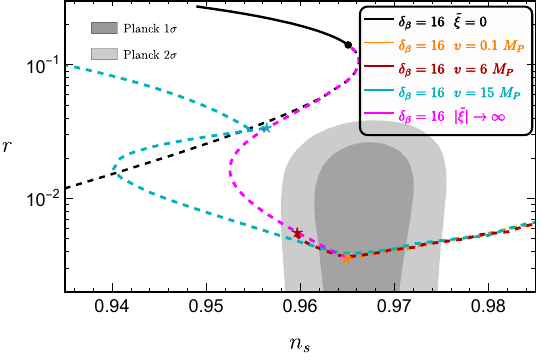}\,
\includegraphics[width=0.49\textwidth]{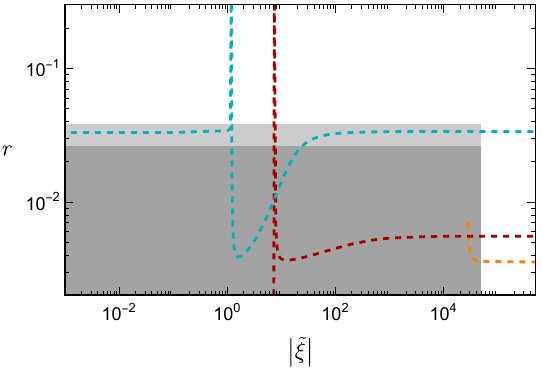}
\\
\includegraphics[width=0.49\textwidth]{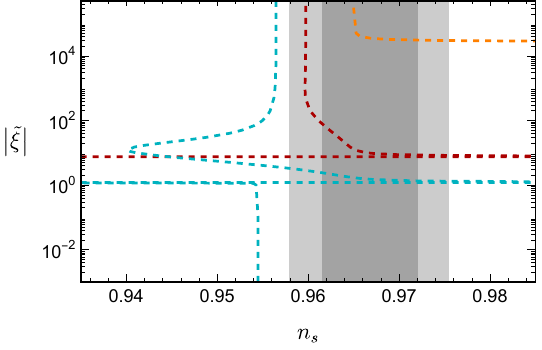}\,
\includegraphics[width=0.49\textwidth]{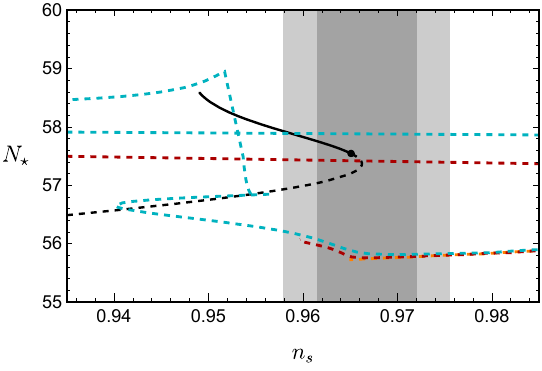}
\\
\includegraphics[width=0.49\textwidth]{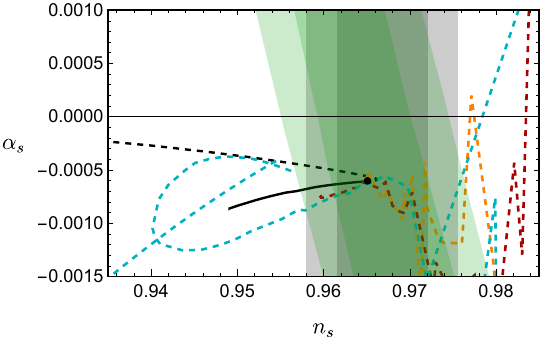}\,
\includegraphics[width=0.49\textwidth]{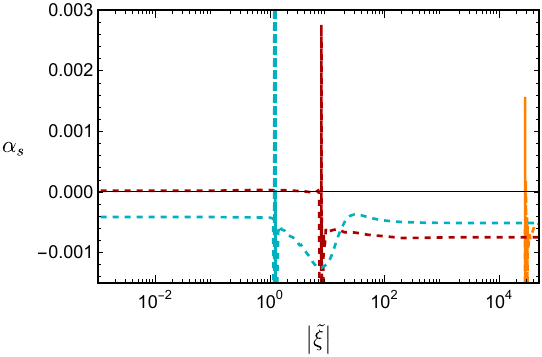}
\caption{Small field inflationary predictions for $\delta_\b=16$ and $\tilde{\xi}<0$. 
$r$ vs.~$n_\text{s}$ (up, left),  $r$ vs.~$|\tilde\xi|$ (up, right),  $|\tilde\xi|$ vs.~$n_\text{s}$ (center, left),  $N_*$ vs.~$n_\text{s}$ (center, right),  $\alpha_\text{s}$ vs.~$n_\text{s}$ (down, left) and $\alpha_\text{s}$ vs.~$|\tilde\xi|$ (down, right) for $N_\star$ given by and $v=0.1M_P$ (orange), $v=6M_P$ (red) and $v=15M_P$ (cyan). The black, pink, gray and green color code are the same as in Fig. \ref{fig:inflation:small:beta0}.}
\label{fig:inflation:small:beta16:neg}
\end{figure}
%%%%%%%%%%%%%%%%%%%%%%%%%FIGURE_END%%%%%%%%%%%%%%%%%%%%%%%%%%%%%%%%%

\begin{figure}[t]
\centering
\includegraphics[width=0.48\textwidth]{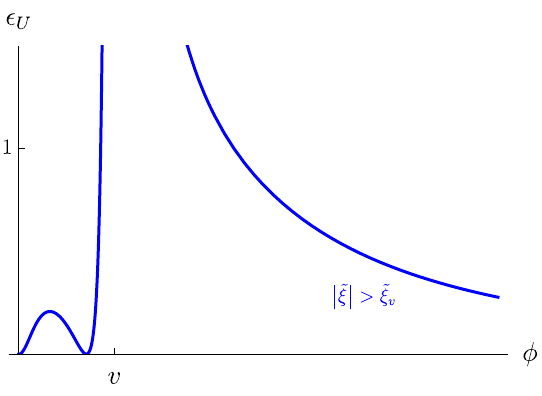}
\includegraphics[width=0.48\textwidth]{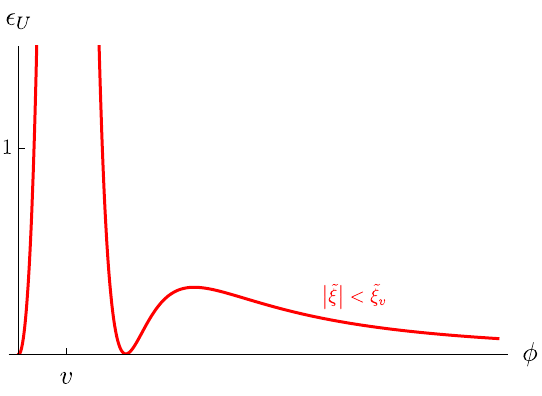}
\caption{$\epsilon_U$ vs. $\phi$ for $\beta(\phi) \neq 0$ with $|\tilde\xi| > \tilde\xi_v $ (blue) and $|\tilde\xi| < \tilde\xi_v $ (red).}
\label{fig:epsplot}
\end{figure}

In this subsection we present the results for the small field inflation scenario. We consider the two reference values\footnote{We have not studied other $\delta_\beta$ values, like $\delta_\beta=1$, because, when $\delta_\beta$ is small, it can be neglected in Eq.~(3.1), making the analysis independent of the sign of the parameter $\tilde{\xi}$. Numerical results show that even for $\delta_\beta = 1$, we remain in the small $\delta_\beta$ regime. In contrast, when $\delta_\beta$ is large (e.g., $\delta_\beta = 16$), the inflationary observables differ significantly depending on whether $\tilde{\xi} > 0$ or $\tilde{\xi} < 0$. } $\delta_\beta=0,16$ respectively in Fig.~\ref{fig:inflation:small:beta0} and Figs.~\ref{fig:inflation:small:beta16:pos} and~\ref{fig:inflation:small:beta16:neg} . In the first case, the system acquires an additional symmetry $\tilde\xi \to -\tilde\xi$, therefore it is enough to just study the $\tilde\xi>0$ case. On the other hand, for the $\delta_\beta=16$ it is needed to study both positive (see Fig.~\ref{fig:inflation:small:beta16:pos}) and negative values (see Fig.~\ref{fig:inflation:small:beta16:neg}) for $\tilde\xi$. Moreover, a zoom out of the results for $r$ vs. $n_s$ when $\delta_\beta=0$ and when $\delta_\beta=16$ and $\tilde\xi>0$ is provided in Fig.~\ref{fig:inflation:small:zoomout}. In all the cases, we considered the reference values\footnote{Since we are considering transPlanckian values for $v$, we also explicitly checked in Appendix~\ref{appendix:masses} that the inflaton mass stays sub-Planckian. On the other hand, the constraint on the amplitude~\eqref{eq:As} ensures that the inflationary energy scale is sub-Planckian when $r$ satisfies the observational bounds~\cite{BICEP:2021xfz}.} $v=0.1 M_P$ (orange), $v=6 M_P$ (red) and $v=15 M_P$ (cyan). Moreover also the predictions for $\tilde\xi=0$ (black) and $|\tilde\xi| \to \infty$ (pink) are shown.

We start commenting the results for $\delta_\beta=0$ shown in Figs.~\ref{fig:inflation:small:beta0} and~\ref{fig:inflation:small:zoomout} by examining the case $v=0.1M_P$.
It is well known that standard SB small field inflation suffers of the $\eta$-problem at small $v$'s. The same happens now.
This can be easily seen by evaluating the second potential slow-roll parameter (see eq.~\eqref{eq:eps:eta}) at $\phi \ll M_P$:
\begin{eqnarray}
    \eta_U(\phi\ll M_P) &=& - 4 \frac{M_P^2}{v^2} \left[1 - \left( 1 + \frac{48\tilde{\xi}^2}{\left(1+4\d_\b^4\right)} \frac{v^2}{M_P^2} \right) \frac{\phi^2}{v^2}  \right] +\mathcal{O}(\phi^4)\,.\label{eq:eta:0_exp:full} 
\end{eqnarray}
It can be proven that the smaller $v$, the smaller is $\phi_*$ (the field value corresponding to $N_*$). Therefore, when $\tilde\xi \to 0$, we have $|\eta_U(0)| \sim  4 \frac{M_P^2}{v^2} > 1$ when $v < 2 M_P$. Indeed, the line for $\tilde\xi=0$ for $v=0.1 M_P$ in Fig.~\ref{fig:inflation:small:zoomout} is not visible because it is out of the validity of the slow-roll approximation. On the other hand, with $\tilde\xi$ increasing it is possible to restore the validity of the slow-roll approximation and even get predictions in the allowed region by the latest combination of Planck, BICEP/Keck and BAO data~\cite{BICEP:2021xfz}. However, since $|\eta_U(0)| \gg 1$ in order to get $n_s$ in the allowed region, $|\eta_U|$ needs to decrease very fast by increasing $|\tilde\xi|$, implying a very big running of the spectral index $\alpha_s$ which is ruled out by Planck legacy data \cite{Planck:2018vyg}. All of this happens in a very small range of big $\tilde\xi$. Then, after $n_s$ reaches its maximum values and turns towards its $\tilde\xi \to \infty$ limit, the running is reduced and the predictions for $r$, $n_s$ and $\alpha_s$ are all within the allowed regions. Therefore agreement with data at sub-Planckian $v$ is achieved at the price of big $\tilde\xi \sim 10^3$.
The predictions for $v = 6 M_P$ have a similar behaviour, with the only difference that the $\eta$-problem is now absent and the $\tilde\xi=0$ is visible in Fig.~\ref{fig:inflation:small:zoomout}. The predictions for $v = 15 M_P$ also follow a similar pattern, with $n_s$ reaching a maximum value in between the $\tilde\xi=0$ and $\tilde\xi \to \infty$ lines. Since the separation between these two limits is much smaller than the one of the previous cases, the running of the spectral index is much more contained. However the predictions for $r$ vs. $n_s$ fall out of the allowed region.

The pattern of the $r$ vs $n_s$ predictions for $\delta_\beta =16$ and $\tilde\xi>0$ (see Fig.~\ref{fig:inflation:small:beta16:pos}) is similar to $\delta_\beta=0$ case, with the predictions moving from a low $n_s$ (large $|\alpha_s|$) at $\tilde\xi=0$ to the allowed region at $\tilde\xi \to \infty$ for $v=0.1,6 M_P$. However, now, no maximum value for $n_s$ is reached at an intermediate $\tilde\xi$ value before the $\tilde\xi \to \infty$ limit (see Fig.~\ref{fig:inflation:small:zoomout}).  The predictions for $v=15 M_P$ remain ruled out.

Last, we discuss the results for $\delta_\beta =16$ and $\tilde\xi<0$ (see Fig.~\ref{fig:inflation:small:beta16:neg}). In this case it is helpful to consider the plot for $\epsilon_U$ vs. $\phi$ given in Fig.~\ref{fig:epsplot}. The relevant feature is the appearance of a local minimum and a local minimum, respectively in $\phi_\text{min, $\epsilon$}$ and $\phi_\text{max, $\epsilon$}$, around $\phimaxk$. When $|\tilde \xi|$ is very small, the peak of the kinetic function is actually in the large field region, therefore the effect is irrelevant and  the results are aligned with the ones of SBI. Then, $|\tilde \xi|$ increases until $|\tilde \xi| > \tilde \xi_v$ and $\phi_\text{max, $k$} < v$. In such a case, $\epsilon_U$ develops the two stationary points mentioned before and inflation mainly happens in the valley between the maximum at $\phi=\phimaxe$ and the pole at $\phi=v$. $r$ increases and $n_s$ decreases, because, even though most of the $e$-folds are done in the valley, in order to get the exact amount,  we need $\phi_*<\phimaxe$.  Then, $|\tilde \xi|$ keeps increasing and the valley gets larger and the peak of the kinetic function keeps moving towards the origin. More and more $e$-folds are done in the valley and $\phi_*$ gets closer to $\phimaxe$. After this, $r$ ($n_s$) reaches a maximum (minimum) value when $\phi_*$ = $ \phimaxe$. Then we reach the configuration $\phi_*  > \phimaxe$ and the valley keeps getting larger. Both $\epsilon_U$ and $\eta_U$ are getting smaller but the first one faster. Therefore $r$ starts decreasing while $n_s$ increasing. Note that all these different regimes take place in a very small range in $\tilde\xi$ and when $\phimaxe$ is actually very close to $v$. Afterwards, by increasing $|\tilde \xi|$ we reach a point where the decreasing of $\eta_U$ is strong enough to induce also $n_s$ decreasing. When $n_s <1$ then $\phi_* > \phimine$. Then the predictions proceed approaching the $|\tilde \xi| \to \infty$ limit and the peak of the kinetic function gets closer to the origin.

We conclude the small field inflation analysis discussing indeed its $|\tilde\xi| \to \infty$ limit. From eq.~\eqref{eq:phi:max} we can see that in such a limit $\phimaxk \to 0$ and $k(\phimaxk) \to \infty$. Even though it is not possible to derive analytical inflationary results for any $v$, it is still possible to derive an explicit limit  for $v \ll M_P$. In such a case we have $\phi_* \ll v \ll M_P$, and the kinetic function is quite well approximated by the pole function
\be
 k(\phi) \approx \frac{6 M_P^2}{ \phi^2} \,.
\ee
From this, we van see that there is no longer any dependence on $\tilde\xi$ or $\delta_\beta$ and the results converge at the leading order to the ones of Starobinsky inflation~\cite{Starobinsky1980} (as confirmed by Figs.~\ref{fig:inflation:small:beta0},~\ref{fig:inflation:small:beta16:pos} and~\ref{fig:inflation:small:beta16:neg}).
%

%%%%%%%%%%%%%%%%%%%%%%%%%%FIGURE%%%%%%%%%%%%%%%%%%%%%%%%%%%%%%%%
\begin{figure}[t]
\centering
\includegraphics[width=0.49\textwidth]{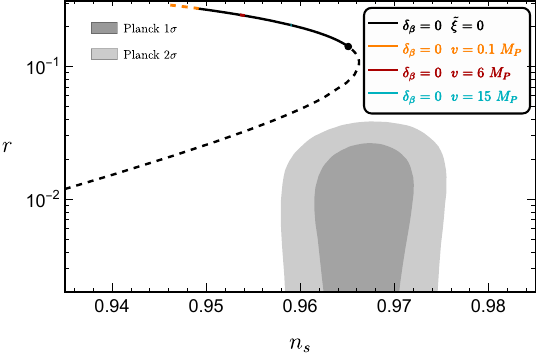}\,
\includegraphics[width=0.49\textwidth]{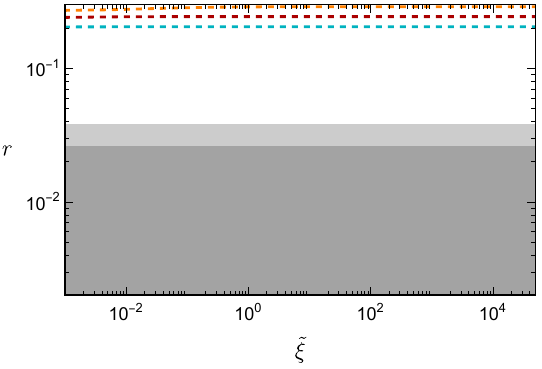}
\\
\includegraphics[width=0.49\textwidth]{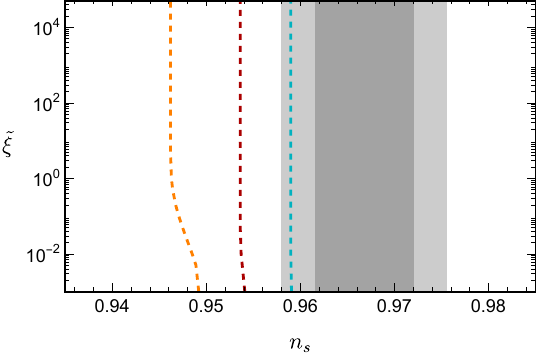}\,
\includegraphics[width=0.49\textwidth]{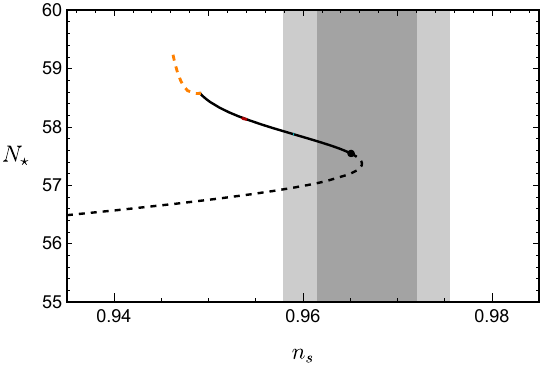}
\\
\includegraphics[width=0.49\textwidth]{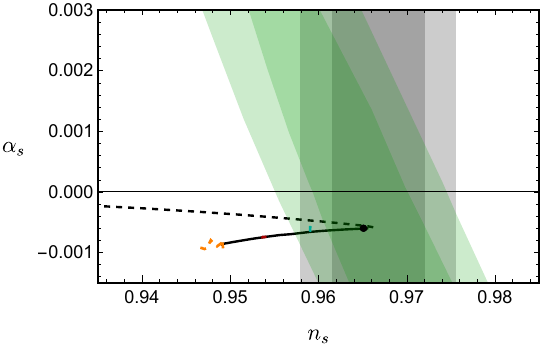}\,
\includegraphics[width=0.49\textwidth]{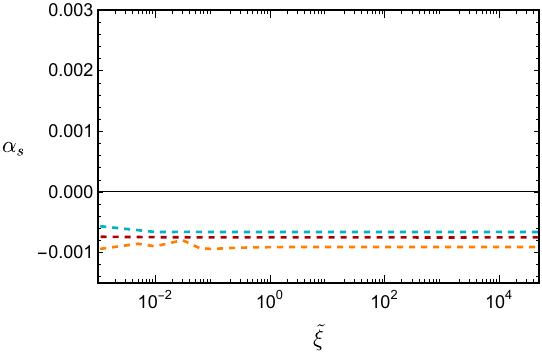}
\caption{Large field inflationary predictions for $\delta_\b=0$.
$r$ vs.~$n_\text{s}$ (up, left),  $r$ vs.~$|\tilde\xi|$ (up, right),  $|\tilde\xi|$ vs.~$n_\text{s}$ (center, left),  $N_*$ vs.~$n_\text{s}$ (center, right),  $\alpha_\text{s}$ vs.~$n_\text{s}$ (down, left) and $\alpha_\text{s}$ vs.~$|\tilde\xi|$ (down, right) for $N_\star$ given by~\eqref{eq:efolds} and $v=0.1 M_P$ (orange), $v=6 M_P$ (red) and $v=15 M_P$ (cyan). For reference, the predictions for $\tilde\xi=0$ (black). The  gray and green color code are the same as in Fig. \ref{fig:inflation:small:beta0}.}
\label{fig:inflation:large:beta0}
\end{figure}
%%%%%%%%%%%%%%%%%%%%%%%%%FIGURE_END%%%%%%%%%%%%%%%%%%%%%%%%%%%%%%%%%

%%%%%%%%%%%%%%%%%%%%%%%%%%FIGURE%%%%%%%%%%%%%%%%%%%%%%%%%%%%%%%%
\begin{figure}[t]
\centering
\includegraphics[width=0.49\textwidth]{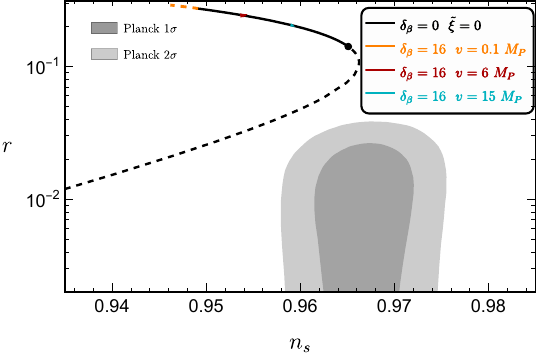}\,
\includegraphics[width=0.49\textwidth]{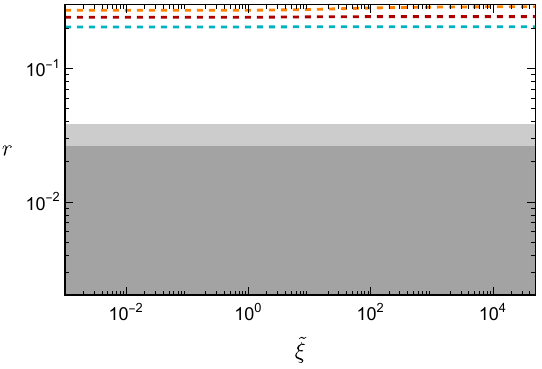}
\\
\includegraphics[width=0.49\textwidth]{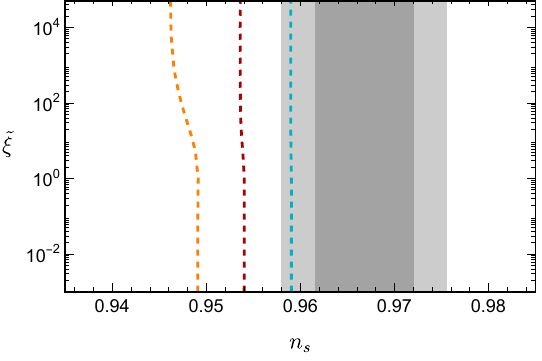}\,
\includegraphics[width=0.49\textwidth]{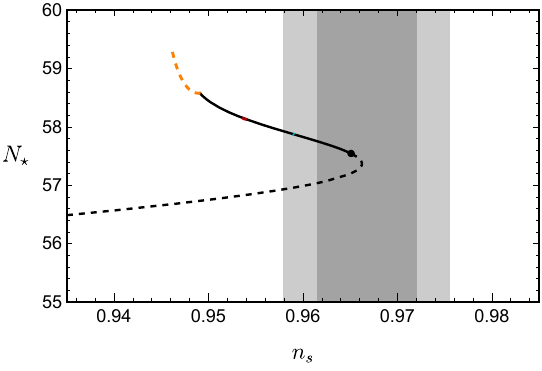}
\\
\includegraphics[width=0.49\textwidth]{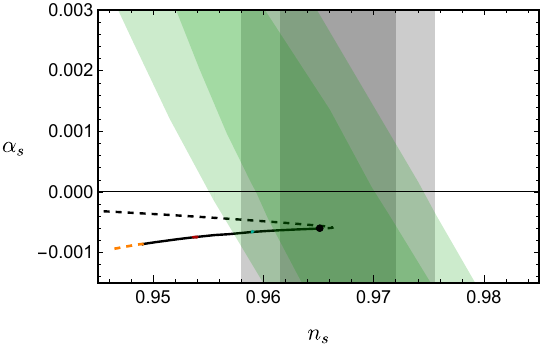}\,
\includegraphics[width=0.49\textwidth]{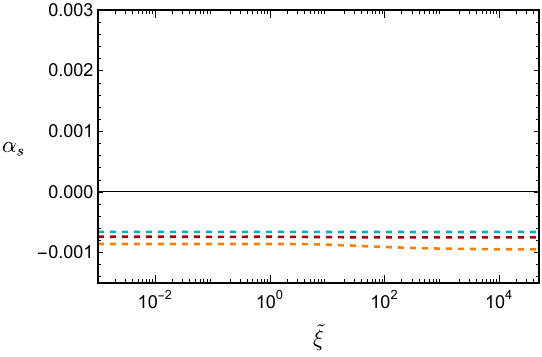}
\caption{Large field inflationary predictions for  $\delta_\b=16$ and $\tilde{\xi}>0$.
$r$ vs.~$n_\text{s}$ (up, left),  $r$ vs.~$|\tilde\xi|$ (up, right),  $|\tilde\xi|$ vs.~$n_\text{s}$ (center, left),  $N_*$ vs.~$n_\text{s}$ (center, right),  $\alpha_\text{s}$ vs.~$n_\text{s}$ (down, left) and $\alpha_\text{s}$ vs.~$|\tilde\xi|$ (down, right) for $N_\star$ given by~\eqref{eq:efolds} and $v=0.1 M_P$ (orange), $v=6 M_P$ (red) and $v=15 M_P$ (cyan). For reference, the predictions for $\tilde\xi=0$ (black). The gray and green color code are the same as in Fig. \ref{fig:inflation:small:beta0}.}
\label{fig:inflation:large:beta16:pos}
\end{figure}
%%%%%%%%%%%%%%%%%%%%%%%%%FIGURE_END%%%%%%%%%%%%%%%%%%%%%%%%%%%%%%%%%

%%%%%%%%%%%%%%%%%%%%%%%%%%FIGURE%%%%%%%%%%%%%%%%%%%%%%%%%%%%%%%%
\begin{figure}[t]
\centering
\includegraphics[width=0.49\textwidth]{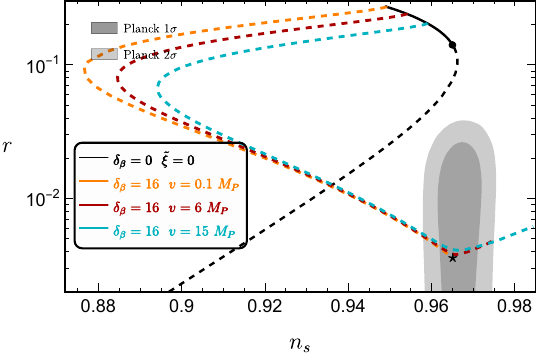}\,
\includegraphics[width=0.49\textwidth]{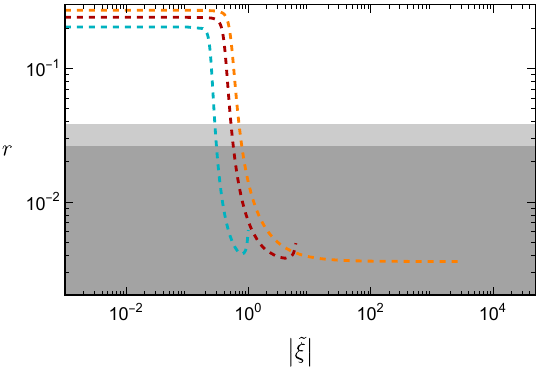}
\\
\includegraphics[width=0.49\textwidth]{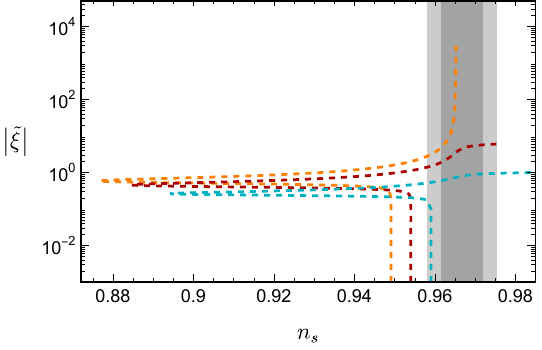}\,
\includegraphics[width=0.49\textwidth]{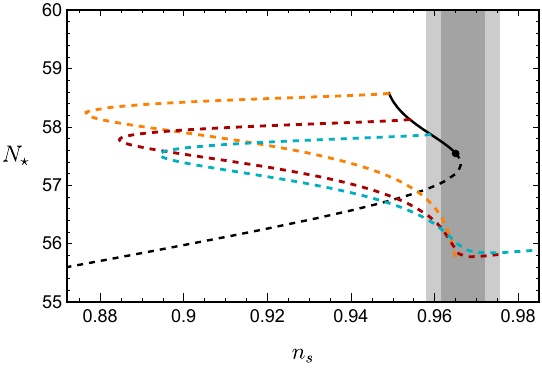}
\\
\includegraphics[width=0.49\textwidth]{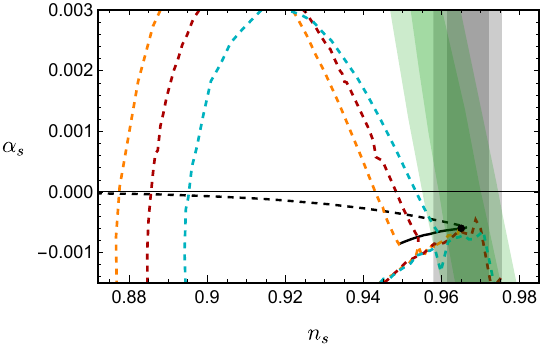}\,
\includegraphics[width=0.49\textwidth]{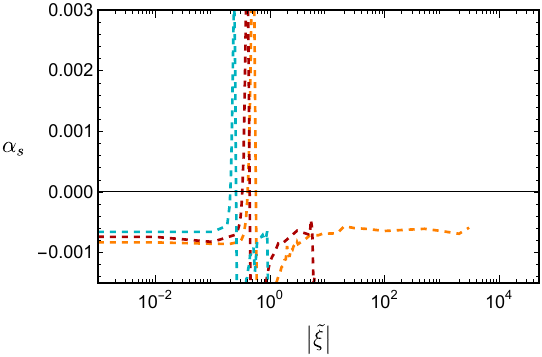}
\caption{Large field inflationary predictions for  $\delta_\b=16$ and $\tilde{\xi}<0$.
$r$ vs.~$n_\text{s}$ (up, left),  $r$ vs.~$|\tilde\xi|$ (up, right),  $|\tilde\xi|$ vs.~$n_\text{s}$ (center, left),  $N_*$ vs.~$n_\text{s}$ (center, right),  $\alpha_\text{s}$ vs.~$n_\text{s}$ (down, left) and $\alpha_\text{s}$ vs.~$|\tilde\xi|$ (down, right) for $N_\star$ given by~\eqref{eq:efolds} and $v=0.1 M_P$ (orange), $v=6 M_P$ (red) and $v=15 M_P$ (cyan). For reference, the predictions for $\tilde\xi=0$ (black). The gray and green color code are the same as in Fig. \ref{fig:inflation:small:beta0}.}
\label{fig:inflation:large:beta16:neg}
\end{figure}
%%%%%%%%%%%%%%%%%%%%%%%%%FIGURE_END%%%%%%%%%%%%%%%%%%%%%%%%%%%%%%%%%

\subsection{Large field inflation}

In this subsection we present the results for the large field inflation scenario. As before, we consider the two reference values $\delta_\beta=0,16$ respectively in Fig.~\ref{fig:inflation:large:beta0} and Figs.~\ref{fig:inflation:large:beta16:pos} and~\ref{fig:inflation:large:beta16:neg} . Again, in the first case, the system acquires an additional symmetry $\tilde\xi \to -\tilde\xi$, therefore it is enough to just study the $\tilde\xi>0$ case. On the other hand, for the $\delta_\beta=16$ it is needed to study both positive (see Fig.~\ref{fig:inflation:large:beta16:pos}) and negative values (see Fig.~\ref{fig:inflation:large:beta16:neg}) for $\tilde\xi$. As before, in all the cases, we considered the reference values $v=0.1 M_P$ (orange), $v=6 M_P$ (red) and $v=15 M_P$ (cyan). Moreover also the predictions for $\tilde\xi=0$ (black) are shown.

We start by discussing the $\delta_\beta \geq 0$ and $\tilde\xi>0$ cases shown in Figs.~\ref{fig:inflation:large:beta0} and~\ref{fig:inflation:large:beta16:pos}. In both cases the predictions of standard SBI are mainly unaffected by the presence of the non-minimal coupling $\beta(\phi)$, because the effect of the non-minimal kinetic function $k(\phi)$ is suppressed. When $\tilde\xi>\tilde\xi_v$, the inflection point is realized in the small field region. On the other hand, when $\tilde\xi<\tilde\xi_v$, $k(\phi)$ is not that peaked either because $\tilde\xi$ is too small, if $\delta_\beta=0$, or because the contributions coming from $\delta_\beta$ cancel each other in eq.~\eqref{eq:phi:max}, if  $\tilde\xi>0$. Therefore such a scenario does not improve the results of standard SBI.

Let us comment now the results for $\delta_\beta = 16$ and $\tilde\xi<0$, given in Fig.~\ref{fig:inflation:large:beta16:neg}. The general behaviour of the results is quite similar to the one seen in~\cite{Racioppi:2024pno}. For all the considered values of $v$, with $|\tilde\xi|$ increasing, first both $r$ and $n_s$ decrease, until $n_s$ reaches a minimum value (dependent on the actual value of $v$). Then $n_s$ starts increasing while $r$ keeps decreasing until it reaches its minimum value (which gets closer to the one corresponding to Starobinsky inflation, with $v$ getting smaller). Then, the results depart away from the Starobinsky region when $|\tilde\xi| > \tilde\xi_v$.

To conclude we note that, since for $|\tilde{\xi}|\phi^2 \gg M_P^2, \phi \gg M_P$ we get  $k(\phi) =1$ and $\phi_\text{max, $k$} \to 0 < v$, the inflationary predictions at $|\tilde\xi| \to \infty$ will ultimately circle back to the $\tilde\xi=0$ independently on $\beta$ or the sign of $\tilde\xi$ when $k(v) \simeq 1$. This happens approximately when $v \gtrsim 10 M_P.$ For numerical convenience, we have never explored such a region, as anyhow ruled out by data.

\section{Conclusions}
\label{sec:concl}
We revisited symmetry breaking inflation in the framework of metric-affine gravity, where the inflaton, featuring a sombrero-hat potential, exhibits a non-minimal coupling function $\beta(\phi)=\delta_\beta^2 + \tilde\xi \frac{\phi^2}{M_P^2}$ with the Holst invariant $\tilde {\cal R}$. Such a non-minimal function induces an inflection point in the inflaton potential. According to the value of $\tilde\xi$, such inflection point will take place before or after the inflaton vev $v$ and improve accordingly the predictions for small field inflation or large field inflation. We performed an explicit numerical analysis with $v=0.1 M_P, 6 M_P, 15 M_P$, $\delta_\beta=0,16$ and both positive and negative values of $\tilde\xi$ for both the small field and the large field inflationary scenarios. We found that it is possible to be in agreement with the latest observational constraints for different values of the considered parameters in both the small and large field regime. Even though very different a priori, both regimes can approach the Starobinsky inflationary predictions for very big $|\tilde\xi|$ and very small $v$. In particular, it is remarkable that the small field scenario becomes again compatible with data even when $v$ is sub-Planckian and $\delta_\beta=0$ if $\tilde\xi$ is big enough. The future experiments with a precision of $\Delta r \sim 10^{-3}$, such as Simons Observatory~\cite{SimonsObservatory:2018koc}, CMB-S4~\cite{Abazajian:2019eic} and LITEBIRD~\cite{LiteBIRD:2020khw}, will be capable test our scenario, specially for the configurations away from the Starobinsky limit.

%-------------------------------------------------------------------------------
\acknowledgments
%-------------------------------------------------------------------------------

This work was supported by the Estonian Research Council grants MOB3JD1202, PRG1055, RVTT3,  RVTT7, and by the CoE program TK202 ``Foundations of the Universe''.

\appendix

\section{Inflaton masses} \label{appendix:masses}

\begin{figure}[t]
\centering
\includegraphics[width=0.49\textwidth]{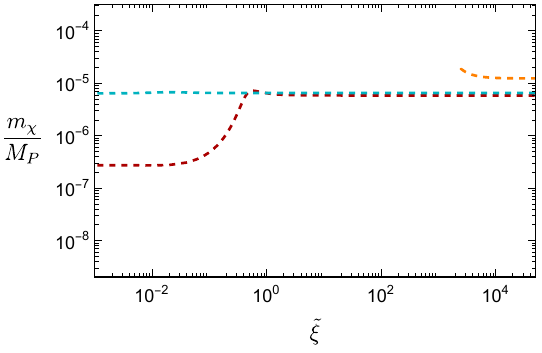}
\includegraphics[width=0.49\textwidth]{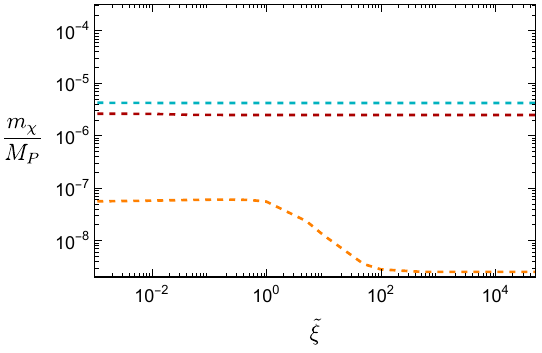}\\
\includegraphics[width=0.49\textwidth]{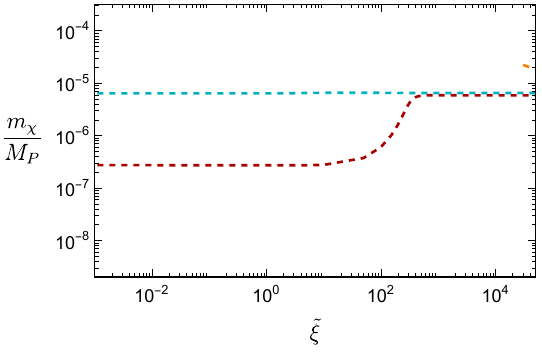}
\includegraphics[width=0.49\textwidth]{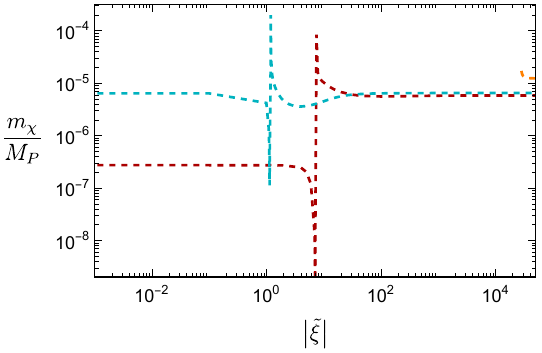}\\
\includegraphics[width=0.49\textwidth]{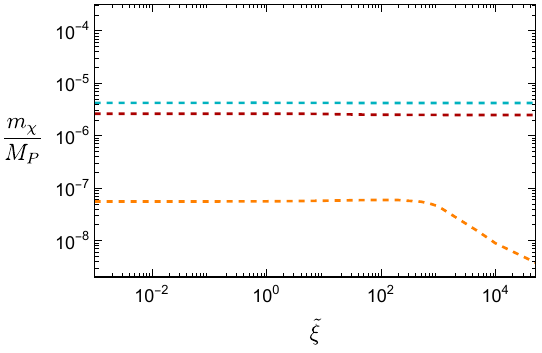}
\includegraphics[width=0.49\textwidth]{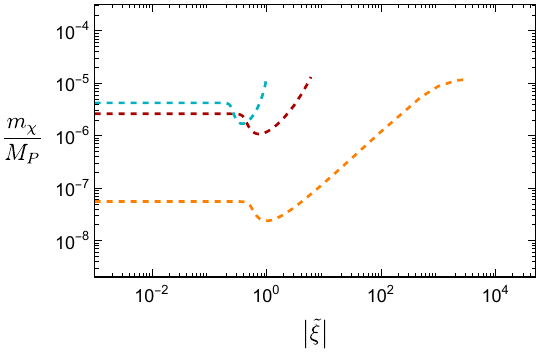}
\caption{Inflaton masses for small field inflation (left column) and large field inflation (right column) respectively for $\delta_\beta=0$ (up), $\delta_\beta=16$ and $\tilde\xi >0 $ (center) and $\delta_\beta=16$ and $\tilde\xi <0 $ (down).}
\label{fig:masses}
\end{figure}

The mass of the canonical inflaton at the minimum is given by
\be
 m_\chi^2 = \frac{2 \lambda  v^2}{1+\frac{24 \tilde{\xi }^2 \delta _v^2}{1+4 \left(\tilde{\xi } \delta _v^2+\delta _{\beta }^2\right)^2}}\label{eq:m2}
\ee
where $\delta_v=\frac{v}{M_P}$
and it is always sub-Planckian for all the studied cases as shown in Figure~\ref{fig:masses}.

\bibliography{SB_MAG}{}
\end{document}